# Near - Infrared Super Resolution Imaging with Metallic Nanoshell Particle Chain Array

Weijie Kong, Xiaoping Zhang,[*] Penfei Cao, Li Gong, Lin Cheng, Xining Zhao, and Lili Yang

*School of Information Science and Engineering, Lanzhou University, Lanzhou 730000, China*

[*]*Corresponding author: zxp@lzu.edu.cn*

**Abstract:** We propose a near-infrared super resolution imaging system without a lens or a mirror but with an array of metallic nanoshell particle chain. The imaging array can plasmonically transfer the near-field components of dipole sources in the incoherent and coherent manners and the super resolution images can be reconstructed in the output plane. By tunning the parameters of the metallic nanoshell particle, the plasmon resonance band of the isolate nanoshell particle red-shifts to the near-infrared region. The near-infrared super resolution images are obtained subsequently. We calculate the field intensity distribution at the different planes of imaging process using the finite element method and find that the array has super resolution imaging capability at near-infrared wavelengths. We also show that the image formation highly depends on the coherence of the dipole sources and the image-array distance.

## 1. Introduction

The resolution of almost all conventional optical system is indispensable governed by the diffraction limit. This resolution limit can be overcome by recovering the evanescent waves in the near-field region. The concept of "perfect lens" in which the evanescent waves can be amplified was proposed firstly by Pendry in 2000 [1]. Since then, superlens [2] and hyperlens [3] made of metamaterials have been widely investigated for their subwavelength imaging and super resolution capabilities. The superlens magnifies the evanescent waves and the subwavelength image can be restored in the near-field [4, 5]. By adding a coupling element to the superlens, the enhanced evanescent waves can be coupled into propagating waves and the super resolution image can be reconstructed in the far-field by collecting the far-field signals from such a lens [6-8]. The hyperlens converts portion of evanescent waves into propagating waves which can be processed by conventional optical system, and the subwavelength featured object can be magnified above the diffraction limit at the output [9, 10]. Although the superlens and hyperlens can image in resolution beyond the diffraction limit, they have disadvantage of one-dimensional imaging only actually [11]. Another kind of diffraction-beating devices based on plasmonics include metallic nanorod array [12, 13] and metallic wire array [14]. The devices can confine and propagate electromagnetic fields on the subwavelength spatial scale. The two-dimensional super resolution image can be obtained with the style of point-to-point. However, the devices did not realize the super resolution imaging in the near-infrared region, which limits the relative applications in biomedical field.

Light confinement below the optical diffraction limit is highly required in nano-optics. Metallic nanoparticles are particularly attractive because they provide a strong three-dimensional subwavelength confinement of optical frequency electromagnetic fields in resonant plasmon modes [15, 16]. The plasmonic waveguides consisting of chains of metallic nanoparticles can support electromagnetic energy transfer below the diffraction limit by means of coupled plasmon modes [17-19]. Afterwards, the metallic nanoparticle chain array for super resolution imaging was proposed [20]. This array has subwavelength resolution capability at visible wavelengths. But the near-infrared super resolution images can not be reconstructed because of the difficulty of tunability. Metallic nanoshell particles are a new class of nanoparticles with highly tunable optical properties [21, 22]. They consist of a dielectric core nanoparticle surrounded by an ultra-thin metal shell. Therefore, the use of nanoshell particles provide additional degrees of freedom for the design of chain waveguides compared to their solid metallic counterparts [23, 24].

In this paper, we propose the metallic nanoshell particle chain array for near-infrared super resolution imaging. This array is composed of many identical metallic nanoshell particle chains. Each chain transports only a single bit of information by coupling the confined resonance modes of localized surface plasmons (LSPs). The whole array transports all discretized bits of information of the object field synchronously and the super resolution images can be reconstructed in the output plane. By tunning the parameters of the metallic nanoshell particle, the plasmon resonance band of the isolate nanoshell particle red-shifts from the visible region of the solid metallic nanoparticle to the near-infrared region. Tunability of the plasmon resonance in the near-infrared region makes the nanoshell particle chain array very useful for non-invasive biomedical super resolution imaging and optical sensing because biological tissue shows maximum optical transmissivity in the "optical window" from 650 to 900 nm [25]. The incoherent and coherent super resolution imaging are all investigated. The simulation results indicate that the gold nanoshell particle chain array can realize the near-infrared super resolution imaging in the incoherent and coherent manners. Compared with the coherent imaging, the incoherent imaging can obtained the higher spatial resolution.

## 2. Imaging Model

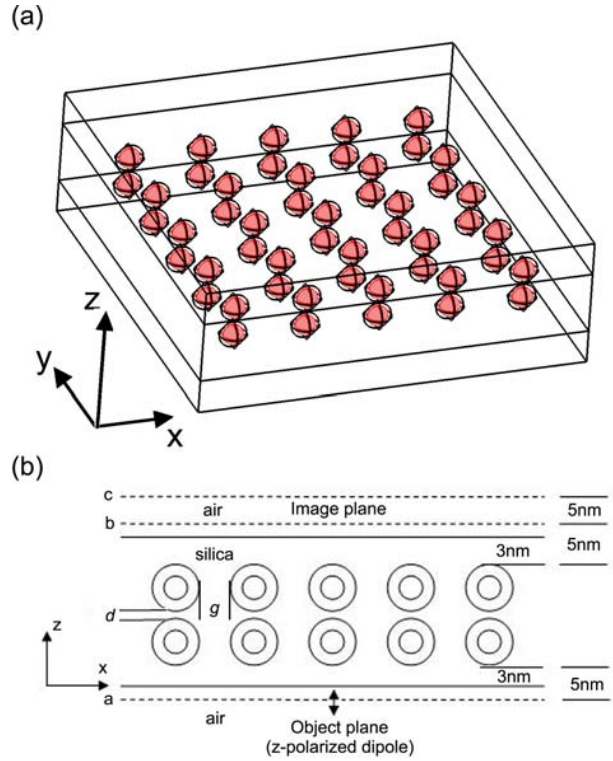

Figure 1. (a) The overall schematic of the metallic nanoshell particle chain array. (b) The corresponding sectional drawing in the x-z plane.

Figure 1 shows a schematic of the studied model, which is made up of metallic nanoshell particle chain array surrounding by silica with refractive index of 1.50. This array is arranged in a square lattice where the interparticle distance is $g$ in the x and y directions and $d$ in the z direction. For each nanoshell particle, the core radius is $r_1$ and the core material is silica, the shell thickness is $s=r_2-r_1$ while the whole nanoparticle radius is $r_2$, the noble metal of nanoshell is gold. When the subwavelength object is located in the object plane, the object source excites the resonance modes of local plasmons in the first nanoshell particles of each chain, which couple with the resonance modes of local plasmons in the second nanoshell particles of each chain at the interparticle gap, and then to the others nanoshell particles in the same fashion, to finally provide a super resolution image in the image plane. In addition, the chains are arranged at enough distance $g$, which dictates the imaging resolution and also guarantees tiny crosstalk between neighbouring chains. Our investigation are carried out with the Radio Frequency (RF) Module for COMSOL Multiphysics 3.5a, which includes a computational electromagnetic code based on the finite element method (FEM) [26]. Scattering boundary conditions (SBCs) and perfectly matched layers (PMLs) are combined together to properly define the computational window.

In our model, the gold nanoshell thickness are smaller than the electron mean free path, which is 42 nm for gold [16], the dielectric function of bulk gold needs to be corrected for the size-limiting effects that restrict the mean free path of conductive electrons. The imaginary part of bulk gold permittivity can be corrected as [27]:

$$\varepsilon'' = \varepsilon_b'' + \Delta\varepsilon = \varepsilon_b'' + iA_L \frac{\lambda_p}{L_{eff}}(\frac{\upsilon_F}{2\pi c})(\frac{\lambda}{\lambda_p})^2 \qquad (1)$$

where $\lambda_p = 1.384 \times 10^{-7}$ m is the wavelength of plasma oscillations, $\upsilon_F = 1.39 \times 10^6$ m/s and c are the Fermi velocity of electrons and the light velocity in vacuum, respectively. The dimensionless parameter $A_L$ is often assumed to be close unity, $L_{eff}$ is the electron effective mean free path of gold and can be described as $L_{eff} = r_2[\frac{1}{1+x^2} - \frac{x}{2} - \frac{(1-x)(1-x^2)}{4(1+x^2)}\ln\frac{(1-x)}{(1+x)}]$, $x = r_1/r_2$ [28]. The dielectric function of bulk gold with frequency-dependent contributions from interband transitions can be approximated by Drude-Lorentz Model [29]. It is found that the imaginary part of corrected gold permitivity increases monotonously with decreasing the nanoshell thickness.

## 3. Imaging with Metallic Nanoshell Particle Chain Array

Before discussing the super resolution imaging performance of gold nanoshell particle chain array, we need find out the fit parameters of chain array for super resolution imaging in the near-infrared region. In this work, the silica with a refractive index of 1.50 is used as the core and surrounding material. The radius of the whole nanoshell particle is 15 nm. The gold nanoshell thickness is determined by the dipole plasmon resonance wavelength in near-infrared region. Here we apply the numerical method to investigate the plasmon resonance for gold nanoshells of different thicknesses based on the extinction efficiency ($Q_{ext} = C_{ext}/(\pi r_2^2)$) of the particle. Fig. 2 shows the calculated extinction efficiency spectra for 15 nm radius silica core-gold nanoshell particles with varying gold nanoshell thickness. The calculated spectrum for the 15 nm radius solid nanosphere has an extinction efficiency maximum at 536 nm corresponding to its dipole plasmon resonance. As the nanoshell thickness $s$ decreases from 15 nm (solid case) to 2 nm, the dipole plasmon resonance progressively red-shifts. In fact, there is a near-exponential red-shift in the dipole plasmon resonance wavelength with decreasing gold nanoshell thickness [22]. In addition, with decreasing nanoshell thickness, the dipole plasmon band increases. At the smallest nanoshell thickness of 2 nm, the dipole plasmon resonance wavelength is 764 nm, the wavelength bandwidth defined by full width at half maximum (FWHM) is larger than 250 nm and spans the entire "optical window" from 650 to 900 nm. Therefore, the gold nanoshell with thickness of 2 nm is chosen to realize the near-infrared super resolution imaging.

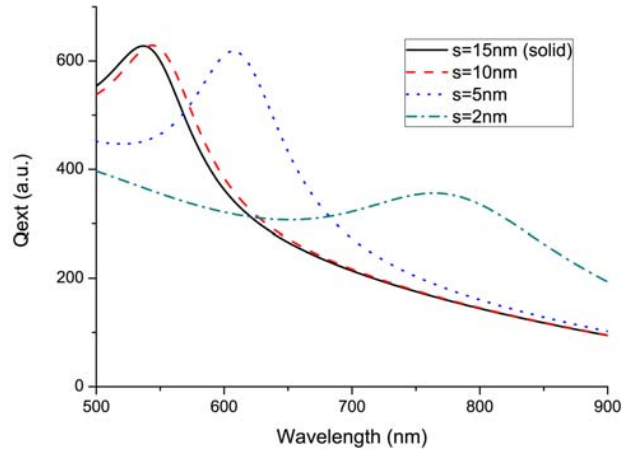

Figure 2. The extinction efficiency spectra of a 15 nm radius silica core-gold nanoshell particle in silica for different shell thickness.

Next, we study the field propagation in single chain consisting of two identical gold nanoshell particles with radius $r_2$ of 15 nm and particles gap $d$ of 2 nm. The gold nanoshell thickness $s$ is 2 nm and the core material is silica. The chain is embedded in silica. A small cylinder with a height of 0.2 nm and a radius of 0.1 nm is positioned in air and 5 nm away from one end of this chain, as shown in Fig. 1(b). The current flowing through the cylinder along the central axis is set to 1 A, which can be regarded as an electric dipole excitation source because of its small dimension. In particular, the field propagation in chains highly depends on the polarization direction of the excitation source. There are two different polarization directions: one parallel to the interparticle axis and the other perpendicular to the axis. In our simulations, we focus on the parallel polarization because the interparticle interactions are much stronger for parallel polarization [30]. Moreover, the chain is driven at the dipole plasmon resonance of an isolate nanoshell particle, i.e., at the incident wavelength of 764 nm. The corresponding corrected gold permittivity is -21.1312+9.5759i. Fig. 3(a) and Fig. 3(b) show the total field intensity distributions ($|E|^2$) in the plane through the chain's axis and that in the cross section normal to the chain's axis through the center of the nanoshell particle near the excitation source. Fig. 3(c) shows the total field intensity profile along the dotted line shown in Fig. 3(b). Fig. 3(a) reveals that a parallel polarization dipole source near the entrance surface of a nanoshell particle excites a local electron oscillation. This oscillation corresponds to the dipole plasmon resonance mode. This mode can be coupled to the neighboring nanoshell particle and then to the next. This so-called domino effect in the plasmon excitation can eventually bring the source information to the other side of the device where the field distribution similar to the source field can be obtained. Fig. 3(b)-3(c) further demonstrate the localization of the plasmon resonance mode. The total field intensity reduces to 1/2 of the intensity near the nanoshell's outer surface when the position is 2.5 nm away from the outer surface of the nanoshell. The corresponding mode dimension is 35 nm. Therefore, the metallic nanoshell particle chain can support the energy propagation with subwavelength confinement of optical frequency electromagnetic fields at plasmon resonance.

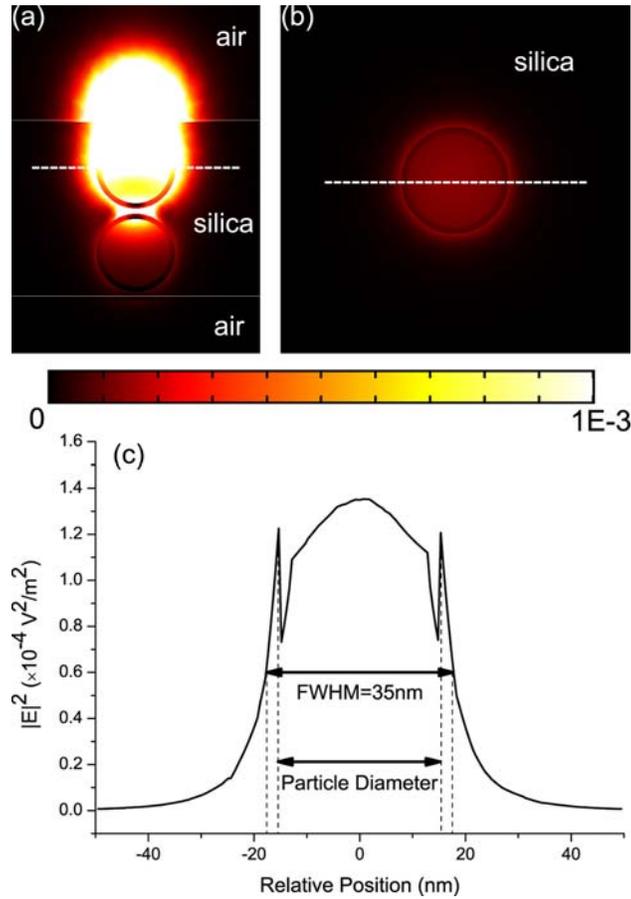

Figure 3. The total field intensity distribution (a) in the plane through the chain's axis and (b) in the cross section normal to the chain's axis through the center of the nanoshell particle near the excitation source. (c) The total field intensity profile along the dotted line shown in Fig. 3(b).

Based on the above results and analyses, the near-infrared super resolution imaging is performed with metallic nanoshell particle chain array. Because the excitation sources include incoherent and coherent light sources, the imaging can be divided into incoherent imaging and coherent imaging. The incoherent imaging is applied to imaging the fluorescent molecules and the coherent imaging is used to laser imaging. In this paper, we investigate not only incoherent imaging but also coherent imaging.

Firstly, we discuss the incoherent super resolution imaging. The imaging array composes of 5×5 gold nanoshell particle chains with the spacing $g$ of 30 nm. Each chain in the array consists of two nanoshell particles with the gap $d$ of 2 nm. The chains are all embedded in silica. Every nanoshell particle has the radius $r_2$ of 15 nm, the nanoshell thickness $s$ of 2 nm and the core material of silica. The point dipoles shaped as the letter 'T' are all polarized parallel and incoherently oscillating. Moreover, the point dipoles are positioned in air to approach the practical case. The chain array is located 2 nm away from the point dipoles, as shown in Fig. 1(b). The calculation volume is 350 nm × 350 nm × 142 nm. The maximum mesh sizes for the cylinder, for the nanoshell particles and for the others regions are 0.05 nm, 10 nm and 15 nm, respectively. The generalized minimal residual method (GMRES) is selected as the linear system solver and the geometric multigrid (GMG) is set as the preconditioner of the solver. Fig. 4(a) shows the total field intensity distribution in the object plane positioned the point dipoles. Fig. 4(b)-4(c) show the intensity distributions in the image plane located 2 nm and 7 nm away from the end of the imaging array, respectively. The intensity profiles crossing the center of the intensity spot (dashed line) are correspondingly illustrated in Fig. 4(d)-4(f). The FWHMs of each spot in the image plane located 2 nm and 7 nm away from the end of the imaging array are 21 nm and 28 nm, respectively. The corresponding ratios of two FWHMs to the wavelength (764 nm) are 1/36 and 1/27, respectively. As can be seen from Fig. 4(b), the letter T is highly resolved in the image plane located 2 nm away from the end of the imaging array. Fig. 4(c) show that spots are blurred and connected with each other with increasing the distance from the end of the imaging array, while the letter T is still clearly resolved. These results demonstrate that the nanopattern is image transferred effectively through a metallic nanoshell particle chain array, containing the super resolution. The spatial resolution is 60 nm (about $\lambda/13$), which is much beyond the diffraction limit.

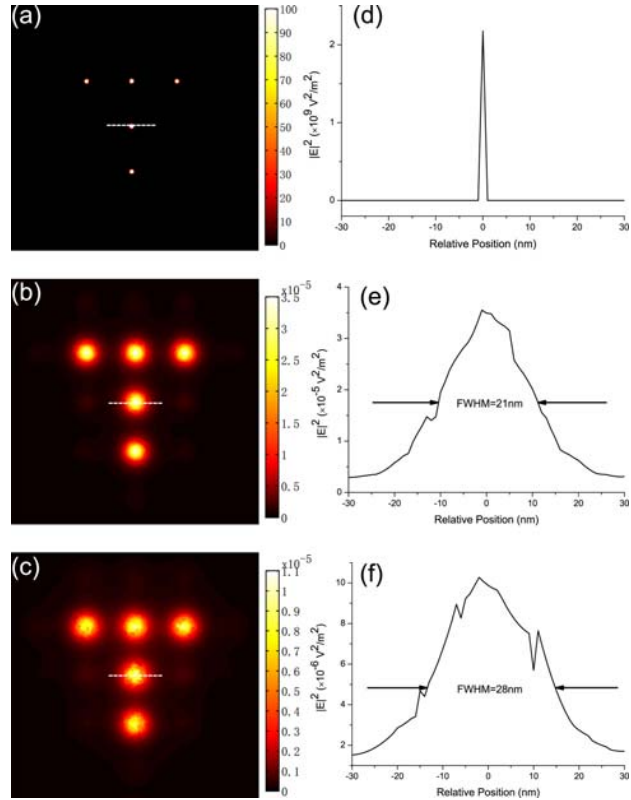

Figure 4. Total field intensity distributions in the incoherent imaging with the chains' spacing of 30 nm (a) in the object plane positioned the dipole sources, (b)-(c) in the image plane located 2 nm and 7 nm away from the end of the imaging array. (d)-(f) The corresponding intensity profiles crossing the center of the intensity spots (dotted lines) shown in Fig. 4(a)-4(c).

Subsequently, the coherent super resolution imaging is developed. The relative parameters keep the same, while the point dipoles are coherently oscillating in phase. Fig. 5(a)-5(c) show the total field intensity distributions in the object plane positioned the dipole sources, in the image plane located 2 nm and 7 nm away from the end of the imaging array, respectively. It is clear that imaging properties are very poor, and the letter T is hardly perceptible. The reason for this phenomenon is related with the strong interferometric effect. Consequently, the spacing between the chains needs to increase to suppress the interactions between them.

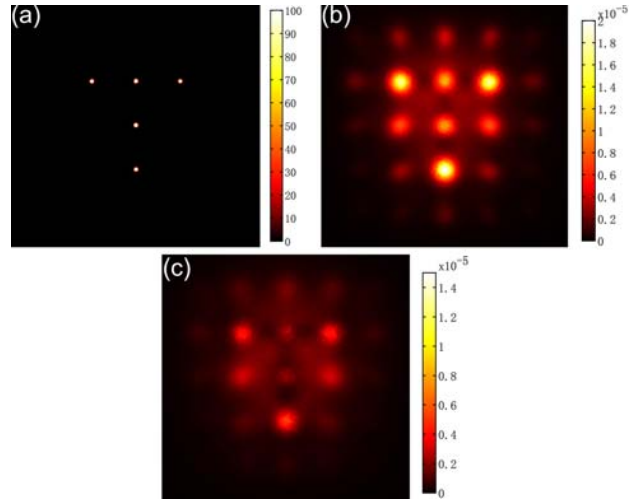

Figure 5. The total field intensity distributions in the coherent imaging with the chains' spacing of 30 nm (a) in the object plane positioned the dipole sources, (b)-(c) in the image plane located 2 nm and 7 nm away from the end of the imaging array, respectively.

In order to reduce the interactions between the chains of the imaging array, the spacing between the chains is changed from 30 nm to 50 nm. The region involved in the calculation is modified to 430 nm × 430 nm × 142 nm. The maximum mesh size for the regions except cylinder and the nanoshell particles are set to 19 nm to reduce memory consumption. The others relative parameters are the same as those in the coherent super resolution imaging simulations with the chains' spacing of 30 nm. The total field intensity distributions in the object plane positioned the dipole sources, in the image plane located 2 nm and 7 nm away from the end of the imaging array are shown in Fig. 6(a)-6(c), respectively. The corresponding intensity profiles crossing the center of the intensity spot (dashed line) are demonstrated in Fig. 6(d)-6(f). The FWHMs of each spot in the image plane located 2 nm and 7 nm away from the end of the imaging array are 21 nm and 26 nm, respectively. The corresponding ratios of two FWHMs to the wavelength (764 nm) are 1/36 and 1/29, respectively. The Fig. 6(b) shows that the super resolution imaging property is excellent. The letter T is well reconstructed in the image plane located 2 nm away from the end of the imaging array. As can be seen from Fig. 6(c), the super resolution imaging property is deteriorated with increasing the distance from the end of the imaging array, but the super resolution is still observed. Therefore, the gold nanoshell particle chain array is able to realize the near-infrared super resolution imaging in the coherent manner. The spatial resolution is 80 nm (about $\lambda/10$), which is much beyond the diffraction limit.

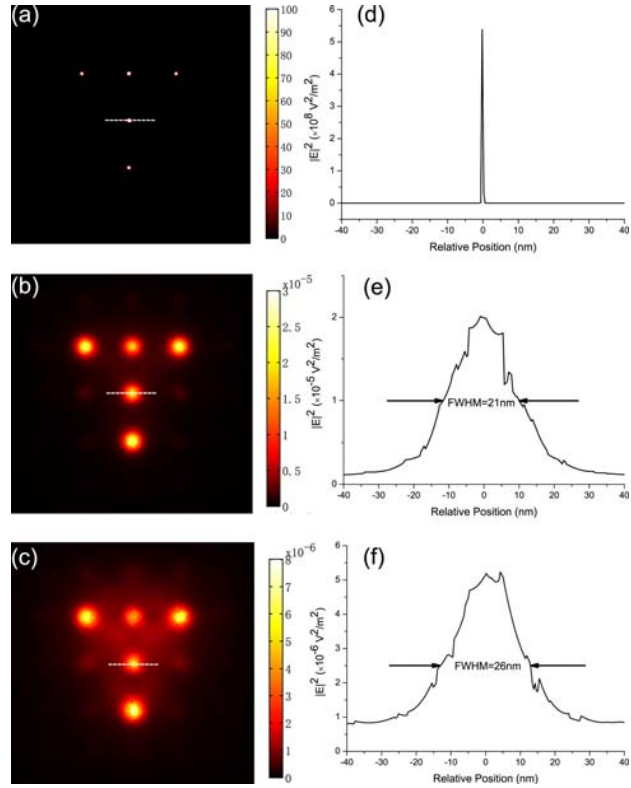

Figure 6. Total field intensity distributions in the coherent imaging with the chains' spacing of 50 nm (a) in the object plane positioned the dipole sources, (b)-(c) in the image plane located 2 nm and 7 nm away from the end of the imaging array. (d)-(f) The corresponding intensity profiles crossing the center of the intensity spots (dotted lines) shown in Fig. 6(a)-6(c).

These results indicate that the gold nanoshell particle chain array has the capability of achieve the near-infrared super resolution image in the incoherent and coherent manners.

Compared with the coherent imaging, the incoherent imaging can obtain the higher spatial resolution.

## 4. Conclusions

In conclusion, the metallic nanoshell particle chain array is numerically confirmed to be a useful device for near-infrared super resolution imaging. The spatial resolution in the case of our used parameters is about λ/13 under the illumination of incoherent light source and about λ/10 under the illumination of coherent light source at the near-infrared wavelength of 764 nm. Compared with the coherent imaging, the incoherent imaging can reach the higher spatial resolution. Our investigation will be useful for non-invasive biomedical super resolution and optical sensing.

## Acknowledgments

This work was supported by the Fundamental Research Funds for the Central Universities of China (lzujbky-2011-k02).